\documentclass[reprint,NumberedRefs]{JASA}

\usepackage{newtxtext}
\usepackage{newtxmath}

\begin{document}

\title[JASA]{Optical Theorem for Measuring the Acoustic Extinction Cross Section of Helmholtz Resonators}

\newcommand{\ITMO}{School of Physics and Engineering, ITMO University,
    197101 St. Petersburg, Russia}
\newcommand{\Qingdao}{Qingdao Innovation and Development Center, Harbin Engineering University, Qingdao 266000, Shandong, China}
\newcommand{\Tongji}{Institute of Acoustics, School of Physics Science and Engineering, Tongji University, Shanghai 200092, China}

\author{Vladimir Igoshin}
\affiliation{\ITMO}
\author{Daniil Klimov}
\affiliation{\ITMO}
\author{Yuri Utkin}
\affiliation{\ITMO}
\author{Sergey Ermakov}
\affiliation{\ITMO}
\author{Mikhail Kuzmin}
\affiliation{\ITMO}
\author{Andrey Bogdanov}
\email{a.bogdanov@hrbeu.edu.cn}
\affiliation{\Qingdao}
\affiliation{\ITMO}
\author{Yong Li}
\affiliation{\Tongji}
\affiliation{\ITMO}
\author{Mihail Petrov}
\email{m.petrov@metalab.ifmo.ru}
\affiliation{\ITMO}

\begin{abstract}
    The optical theorem is a powerful tool of scattering theory that directly relates the extinction cross section of a scatterer to its forward scattering amplitude.
    While widely used in electromagnetism and optics, its application in acoustics has remained limited, primarily due to experimental challenges.
    These include the finite size of practical sound sources and the stringent requirements for detecting weak scattered signals. In this work, we analyze these limitations and develop a robust methodology for measuring the acoustic extinction cross section under realistic conditions, including non-ideal anechoic environments.
    The approach is applied experimentally to a Helmholtz resonator in an imperfect anechoic environment with standing-wave resonances.
    The retrieved extinction cross section exhibits a single pronounced resonance near 2000 Hz.
    The resonance position, peak magnitude, and spectral lineshape are consistent with full-wave numerical simulations.
    The results demonstrate that, when combined with appropriate data processing, the optical theorem provides a simple and reliable tool for characterizing acoustic resonators, opening new opportunities for quantitative analysis of acoustic scattering and absorption phenomena.
\end{abstract}

\maketitle

\section{Introduction\label{sec:1}}

Acoustic resonators play a central role in controlling sound waves and serve as fundamental building blocks for a wide class of acoustic devices~\cite{Fahy2000,Kinsler2000} and acoustic metamaterials~\cite{Liu2000,Ma2016Feb,Cummer2016,Chen2023Jul}.
Among the variety of resonators, Helmholtz resonators are of particular importance due to their simple design, pronounced subwavelength resonance, and broad applicability~\cite{Langfeldt2022Oct,Ingard1953,Zhen2025Jun,Toftul2025Dec}.
They are widely used for noise reduction, acoustic filtering, spectral shaping, and the realization of metasurfaces with engineered wavefront control~\cite{Xie2014Nov,Cheng2015Oct,Cummer2016,Krasikova2024Jan}.
The accurate experimental characterization of their properties, including resonant frequency, linewidth (Q-factor), and energy losses, is therefore essential for both fundamental studies and practical design of acoustic systems.

\begin{figure}[t]
    \includegraphics{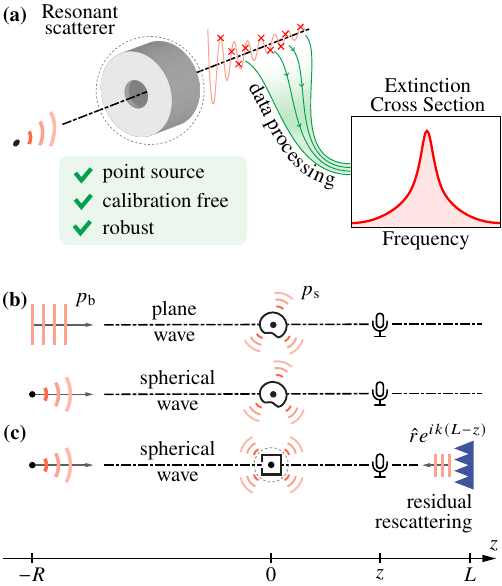}
    \caption{
        \label{fig:lines}
        (a) Concept of the proposed extinction cross section measurement method.
        The acoustic field is measured behind a scatterer and processed to extract the extinction cross section spectrum.
        (b) Geometries of the problem for theoretical investigation for plane wave and spherical wave cases.
        In all geometries, the scatterer is placed at $z=0$, the point source at $z=-R$, and the detector (microphone) at a position on $z$-axis.
        (c) The model geometry of the experiment.
        The rescattered waves from the anechoic chamber walls are taken into account, and an effective distance $L$ from the scatterer to the wall is introduced.
    }
\end{figure}

A universal quantity that fully characterizes the interaction of an open resonator with an incident wave is the {\it extinction cross section} (ECS), a fundamental concept of scattering theory~\cite{Newton,mishchenko2002,Hulst1981}.
It quantifies the total removal of energy from the incident field due to both scattering and absorption~\cite{Newton,mishchenko2002,Rayleigh}.
While separate measurements of scattering and absorption cross sections are generally challenging, especially in three-dimensional geometries, the ECS can be accessed experimentally through interference effects in the forward direction.
Indeed, scattering theory establishes that the extinction cross section is proportional to the imaginary part of the forward scattering amplitude, or equivalently, to the imaginary part of the trace of the scattering (S-) matrix~\cite{mishchenko2002,Marston2001,Newton}.
This relation is known as the {\it optical theorem}~\cite{Newton,gouesbet_failures_2023}.
Its key advantage is that it enables determination of the ECS from measurements in a single direction, significantly simplifying experimental requirements.

The optical theorem is extensively used in electromagnetics and radiophysics, where the forward radar cross section (RCS) serves as a central observable in scattering and antenna theory~\cite{Larsson2013,Knott2004,Richards2005}, and is directly linked to the extinction cross section~\cite{Larsson2013}.
In this context, calibration targets with known RCS, such as metallic spheres, are routinely employed~\cite{Larsson2013,Knott2004}.
High-precision RCS measurements have also enabled investigations of fundamental phenomena in open and non-Hermitian systems~\cite{zhang_non-hermitian_2025}.

In acoustics, a rigorous mathematical analogue of the optical theorem exists and has been extended beyond plane-wave excitation to arbitrary beam profiles using multipole expansions and angular spectrum methods~\cite{Kargl1990Aug,Silva2011,Sapozhnikov2013Feb,Mitri2014Nov,gouesbet_failures_2023}.
Within this framework, the generalized optical theorem provides access to both the imaginary component of the angular scattering amplitude and the extinction cross section for arbitrary beam profiles~\cite{Marston2001,Zhang2019Mar}.
Equivalent formulations based on spherical-wave decompositions lead to the generalized Lorenz-Mie theory, which offers a unified description of scattering from finite sources and structured wavefronts.
This formalism has been widely used to compute acoustic radiation forces and torques, with established applications in particle manipulation, including acoustic levitation, tweezers, and sorting~\cite{Sapozhnikov2013Feb,Zhang2016Aug,Zhang2013Sep,gouesbet_failures_2023}.

Despite the theoretical progress, practical application of the optical theorem in acoustics remains limited, particularly for broadband measurements in realistic three-dimensional environments.
Only a few experimental studies have attempted to measure the ECS in acoustics.
In Ref.~\citen{Sanchis2013Mar}, an indirect approach based on spatial mapping of the acoustic field behind the scatterer was used to infer extinction effects.
A related method was demonstrated in a two-dimensional configuration in a water tank~\cite{Zhang2011Jan}.
Direct measurements based on the optical theorem were proposed in Ref.~\citen{Smith1985Jan}, where experiments in an underwater environment showed agreement with theoretical predictions.
However, these studies were limited either to indirect inference, reduced dimensionality, or measurements at a single frequency point, which prevents full characterization of resonant behavior.

The main difficulty in applying the optical theorem to acoustic experiments lies in the presence of non-ideal conditions.
Realistic sound sources generate spherical rather than plane waves, scatterers exhibit complex multipole responses~\cite{Sapozhnikov2013Feb, smagin_acoustic_2024, tsimokha_acoustic_2022}, and measurement environments introduce parasitic effects such as residual reflections and standing-wave noise~\cite{timankova_experimental_2026}.
These factors violate the assumptions underlying the standard formulation of the optical theorem and lead to significant errors in directly retrieved ECS values.

In this work, we address these limitations and develop a practical framework for measuring the acoustic extinction cross section under realistic experimental conditions. The key details of our approach are illustrated  in  \autoref{fig:lines}.
First, we derive a generalized form of the optical theorem for spherical incident waves and quantify the impact of finite source and detector distances.
Second, we introduce a calibration-free, two-step reconstruction procedure that separates the background field from the scattered contribution and compensates for residual rescattering in imperfect anechoic environments.
Finally, we validate the proposed method experimentally by measuring the ECS spectrum of a Helmholtz resonator and comparing it with numerical simulations.

The main contributions of this work are as follows:
(i) a formulation of the optical theorem applicable to spherical wave excitation relevant for typical acoustic sources;
(ii) a robust and calibration-free methodology for ECS retrieval in non-ideal environments;
and (iii) experimental demonstration of broadband ECS measurements for subwavelength acoustic resonators.
\section{The Optical Theorem for Incident Spherical Wave\label{sec:2}}

\subsection{Optical Theorem for Plane Waves}

It is well-known that the optical theorem is valid for plane wave incidence and breaks for arbitrary incident fields~\cite{krasavin2018,newton1976,mishchenko2002,gouesbet_failures_2023}.
In experiments, however, generating plane-wave-like incident fields requires sophisticated loudspeaker arrays~\cite{Diaz-Rubio2019Feb,Sapozhnikov2013Feb}.
Many real acoustic sources, such as single loudspeakers, behave like point-like sources, since their lateral size may be much smaller than the wavelength.
Even loudspeakers with non-trivial directivity can  be modeled as clusters of point sources~\cite{magalhaes2004sound}.
At sufficiently large distance the field of a point source in the direction of propagation approaches a plane-wave as the radius of curvature of the phase front increases.
However, it still remains unexplored what distance can be considered sufficiently large and how deviations from the plane-wave condition affect measurements of ECS.
Thus, at the first stage, we introduce the ECS expression corrected to account for non-plane wave incidence.
The optical theorem formula should be corrected in the case of spherical incident waves.

We start by considering a scatterer located at the origin of the system.
As the optical theorem requires only fields calculated in the forward direction, we will refer to $z$ as the distance from the origin of the system to the point located on the main axis, as shown in \autoref{fig:lines}~(b).

The forward scattering in the case of an incident plane wave (PW) $p_\text{b}(z) = p_0 e^{ikz}$ in the far-field can be expanded~\cite{mishchenko2002}
\begin{equation}
    p_\text{s}(z) = p_0 \left[\frac{f}{kz}e^{ikz} + \mathcal{O}\left(\frac{1}{(kz)^2}\right)\right],
    \label{eq:p_s_therory_expansion}
\end{equation}
where $f$ is the forward scattering amplitude, which depends only on the incident field frequency $\omega$, $k=\omega / c$ is the wavenumber, and $c$ is the speed of sound.
Here, $\mathcal{O}(1/x)$ denotes the big-O term whose magnitude is bounded by $C/x$ for some constant $C$ as $x\to\infty$.
While there is an ambiguity in choosing the multipole decomposition origin, there are particular optimal ways to do that~\cite{Ustimenko2025Apr}. The optical theorem can be expressed as follows
\begin{equation}
    \sigma_\text{ext} = \frac{4\pi}{k^2}\operatorname{Im} f = \frac{4\pi}{k}\operatorname{Im}\left[\frac{p_\text{s}(z)}{p_0 e^{ikz}}\right]z + \mathcal{O}\left(\frac{1}{k^3z}\right),
\end{equation}
where $\sigma_\text{ext}$ is the ECS~\cite{Hulst1981,Newton}, and here and in the following, $z$ denotes the position of the detector (microphone).

\begin{figure}
    \includegraphics{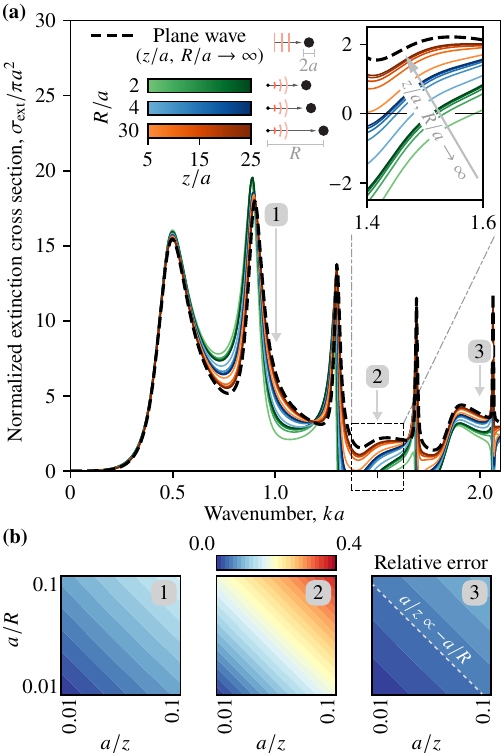}
    \caption{
        \label{fig:opt_t}
        (a)~Normalized extinction cross section calculated with optical theorem for different distances between sphere center and point source $R/a$ and points of measurement $z/a$.
        Extinction cross section in the case of a plane wave is presented by the black dashed line.
        (b)~Dependence of relative error of extinction cross section calculated with optical theorem for different $a/R$ and $a/z$ for $ka=1$, $ka=1.5$, and $ka=2$.
    }
\end{figure}

Both incident and scattered fields can be expanded in terms of spherical wave harmonics $Y_\ell^m(\theta, \varphi) j_\ell(kr)$ and $Y_\ell^m(\theta, \varphi) h^{(1)}_\ell(kr)$~\cite{Mitri2014Nov, Sapozhnikov2013Feb, Ambrosio2024Sep}.
In this basis, one can provide the connection between the scattered and incident amplitudes with the help of T-matrix formalism:
\begin{equation}
    \mathbf{p}_\text{s} = \mathbf{\hat{T}} \mathbf{p}_\text{b},
    \label{eq:T-matrix}
\end{equation}
where $\mathbf{p}_\text{s,b}$ are vectors containing the expansion coefficients for the scattered and incident fields and $\mathbf{\hat{T}}$ is the T-matrix~\cite{Waterman1969Jun, Mitri2014Nov, Sapozhnikov2013Feb}.

\subsection{Extension to Spherical Waves}

The incident field provided by the loudspeaker can be approximated by a point source located at the distance $R$ from the scatterer (see \autoref{fig:lines}~(b)) taking the form of spherical wave (SW):
\begin{equation*}
    p_\text{b}(z) = p_0 kR \dfrac{e^{ikz}}{k(z+R)}.
\end{equation*}
Thus, $p_0\equiv p_\text{b}(0)$ is the incident pressure at the scatterer origin.
The factor $e^{ikR}$ is therefore included in the normalization.
The expansion of $p_\text{b}(z)$ in the forward direction (at $\theta=0, \varphi=0$) in terms of  the spherical wave harmonics in the vicinity of the scatterer $|z|<R$ is given by:
\begin{multline}
    p_\text{b}(z) %
    = p_0 \sum_{\ell=0}^\infty i^\ell(2\ell+1)  j_\ell(kz)\left[\frac{kR}{e^{ikR}}(i)^{\ell+1} h_\ell^{(1)}(kR)\right] \\
    = p_0 \sum_{\ell=0}^\infty i^\ell(2\ell+1)  j_\ell(kz)\left[1 + \frac{i(\ell+1)\ell}{2kR} + \mathcal{O}\left(\frac{1}{(kR)^2}\right)\right],
    \label{eq:p_b_expansion}
\end{multline}
where $j_\ell$ and $h_\ell^{(1)}$ are the spherical Bessel and 1st kind Hankel functions.
A detailed derivation is given in \autoref{app:expansion}.
The terms in square brackets correspond to the beam-shape coefficients of the incident spherical wave~\cite{gouesbet_failures_2023,Zhang2016Aug}.
For comparison, a PW propagating along the $z$ axis has the expansion
\begin{equation*}
    p_\text{b}(z) = p_0 e^{ikz} = p_0\sum_{\ell=0}^\infty i^\ell (2\ell+1)  j_\ell(kz).
\end{equation*}
Therefore, the SW incident coefficients differ from the PW coefficients by terms proportional to powers of $1/(kR)$.
By linearity of the T-matrix relation in \autoref{eq:T-matrix}, the same corrections enter the scattered field.
In the forward far field this gives
\begin{equation*}
    p_\text{s}(z)=p_0\frac{e^{ikz}}{kz}\left[f+\mathcal{O}\left(\frac{1}{kR}+\frac{1}{kz}\right)\right],
\end{equation*}
where $f$ is the forward scattering amplitude for PW incidence.
The intermediate steps are given in \autoref{app:f_expansion}.
Substitution into the plane-wave optical theorem yields
\begin{equation*}
    \sigma_\text{ext} = \frac{4\pi}{k}\operatorname{Im}\left[\frac{p_\text{s}(z)}{p_0 e^{ikz}}\right]z + \mathcal{O}\left(\frac{1}{k^3z} + \frac{1}{k^3R}\right).
\end{equation*}
Finally, noting that $p_0 e^{ikz} = p_\text{b}(z)\cdot[z+R]/R$, one can formulate the optical theorem in the case of spherical waves:
\begin{equation}
    \boxed{\sigma_\text{ext} = \frac{4\pi}{k}\operatorname{Im}\left[\frac{p_\text{s}(z)}{p_\text{b}(z)}\right]\frac{zR}{z+R} + \mathcal{O}\left(\frac{1}{k^3z}+ \frac{1}{k^3R}\right)}.
    \label{eq:opt-theorem-SW}
\end{equation}

The coefficient $R/(z+R)$ tends to 1 as $R\to+\infty$ and the formula becomes the optical theorem for a plane wave.
A similar formula without the error term was originally proposed in Ref.~\citen{Smith1985Jan}.
\autoref{eq:opt-theorem-SW} is the main tool for calculating ECS and contains only measurable quantities and allows retrieving the ECS in the case of a spherical incident wave.
The formula also shows that the accuracy of the directly measured ECS depends both on the position of the SW source $R$ and the position of the detector $z$.
We emphasize that \autoref{eq:opt-theorem-SW} offers accuracy estimates only with respect to the geometric parameters $R$ and $z$; it does not constrain the accuracy in $k$, as the residual terms are determined by the specific scattering physics encoded in the T-matrix.

\subsection{Accuracy Analysis: Mie Problem Example\label{subsec:mie}}

The accuracy of the ECS retrieval in the suggested geometry can be illustrated with a model example of sound scattering from a resonant spherical particle in air.
This is an acoustic analogue of the well-known Mie problem (see, for example, Ref.~\citen{anderson_sound_1950} or Appendix A in Ref.~\citen{Toftul2025Sep}).
We consider a resonant scattering regime when particle and host media properties are related as follows: density $\rho_\text{sphere}/\rho_\text{host} = 1$, speed of sound $c_\text{sphere}/c_\text{host} = 0.3$.
We consider only the pressure wave propagation in the sphere and the medium, in other words, a fluid/gas bubble in a fluid/gas medium.
In \autoref{fig:opt_t}~(a), the ECS of this particle at different wavenumbers $ka$ is shown, where $a$ is the sphere radius.
The spectrum demonstrates pronounced Mie-type  resonances.
The dashed line shows the exact ECS for PW illumination, calculated from \autoref{eq:sigma_ext_sphere}, which can be derived from the rigorous solution~\cite{anderson_sound_1950}.
The colored curves show the ECS for a point source at distance $R$ and detector (microphone) at distance $z$.
The ECS is retrieved from the leading explicit term of \autoref{eq:opt-theorem-SW}, with $p_\text{b}(z)$ and $p_\text{s}(z)$ evaluated analytically for the SW and spherical scatterer, respectively, as detailed in \autoref{app:sphere}.

Although the main resonant features remain visible at relatively small $z/a$ and $R/a$, significant errors occur away from the resonances, sometimes producing negative ECS values.
These negative values do not imply that the particle amplifies the sound, they arise from additional forward amplitude phase contributions introduced by error terms.
As the source and detector distances increase, the SW retrieval converges to the PW reference, as shown in the inset of \autoref{fig:opt_t}~(a).
In \autoref{fig:opt_t}~(b), the relative error, defined as the relative difference between the ECS calculated from \autoref{eq:opt-theorem-SW} and theoretical prediction for PW, is shown for different wavenumbers $ka$.
All plots show agreement with the predicted accuracy $\mathcal{O}\left(1/k^3z + 1/k^3R\right)$.
The suppression of the errors at the resonance can be explained by a single multipole dominance.

The required distance $R$ to approximate an incident plane wave with a point-source field roughly follows from \autoref{eq:p_b_expansion}: one needs $1 \gg \frac{(\ell+1)\ell}{2kR}$ for all relevant $\ell$.
The number of spherical-wave harmonics needed to represent the field on a sphere of radius $a$ is bounded by $\ell_\text{max} \approx ka$~\cite{jensen2004number}.
Therefore, the condition on $R$ can be estimated as $R \gg ka^2$, which is the standard far-field criterion and coincides with the Fraunhofer diffraction condition~\cite{born2013principles}.

\section{Measuring ECS in the experimental setup\label{sec:3}}

\autoref{eq:opt-theorem-SW} explicitly shows that the sound source and detector must be placed far away from the scatterer ($kz \gg 1$, $kR \gg 1$) to obtain ECS close to the theoretical prediction.
However, this also means that the source signal and the scattered signal will be very weak and can be easily hindered by noise.
Indeed, let us assume that measurements are carried out in a room with non-zero residual rescattering from walls, e.g., an imperfect anechoic chamber. Then, our background and scattering pressure have additional contributions
\begin{eqnarray*}
    \widetilde{p}_\text{b} = p_\text{b} + \Delta p_\text{b}, \quad \widetilde{p}_\text{s} = p_\text{s} + \Delta p_\text{s},
\end{eqnarray*}
where the tilde symbols stand for the measured quantities, $\Delta p_\text{b}$ and $\Delta p_\text{s}$ represent noise background and scattered field due to the residual reflections from the walls.
It can be explicitly shown that at large distances, assuming  $|\Delta p_\text{b}| \ll |p_\text{b}|$ (see \autoref{app:sigma_noisy} for the derivation details):
\begin{equation}
    \left|\widetilde{\sigma}_\text{ext} - \sigma_\text{ext}\right| \lesssim \frac{4\pi}{k^2}\left[\frac{|\Delta p_\text{s}|}{|p_0|}kz + |f|\frac{|\Delta p_\text{b}|}{|p_0|}\left(\frac{z}{R}+1\right)\right],
    \label{eq:sigma_noisy_error}
\end{equation}
which means that even small reflections and perturbations in the measurement system result in a linear growth of error in ECS with  the distances $z$ in the far-field, where the  optical theorem should be applied.
Therefore, direct ECS measurements in the far-field region by applying \autoref{eq:opt-theorem-SW} are impossible under such imperfect conditions.

\begin{figure}
    \includegraphics{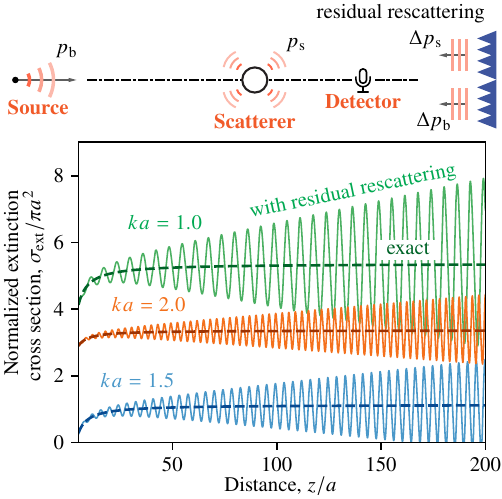}
    \caption{
        \label{fig:noisy_sigma}
        Extinction cross section retrieved from simulations as a function of the detector position $z$ without (dashed line) and with (solid lines) residual rescattering.
        Different colors correspond to different wavenumbers from \autoref{fig:opt_t}.
        The amplitude of residual scattering $\Delta p_b=\Delta p_s=0.002p_0$.
    }
\end{figure}

In order to illustrate this, we used the analytical model based on the Mie scattering problem discussed in \autoref{subsec:mie} to analyze the contribution of noise or residual rescattering to the ECS.
\autoref{fig:noisy_sigma} shows the predicted ECS from a resonant particle considered in \autoref{fig:opt_t} for $R/a=4$ accounting for the additional weak constant residual signal $\Delta p_b=\Delta p_s=0.002 p_0$.
One can see that the expected value of measured ECS in the absence of noise approaches the theoretical prediction and is shown by dashed lines at different frequencies.
At the same time, adding even weak noise results in an oscillating error which grows linearly with the increasing distance from the scatterer to the detector, in full agreement with \autoref{eq:sigma_noisy_error}.

\begin{figure}
    \includegraphics{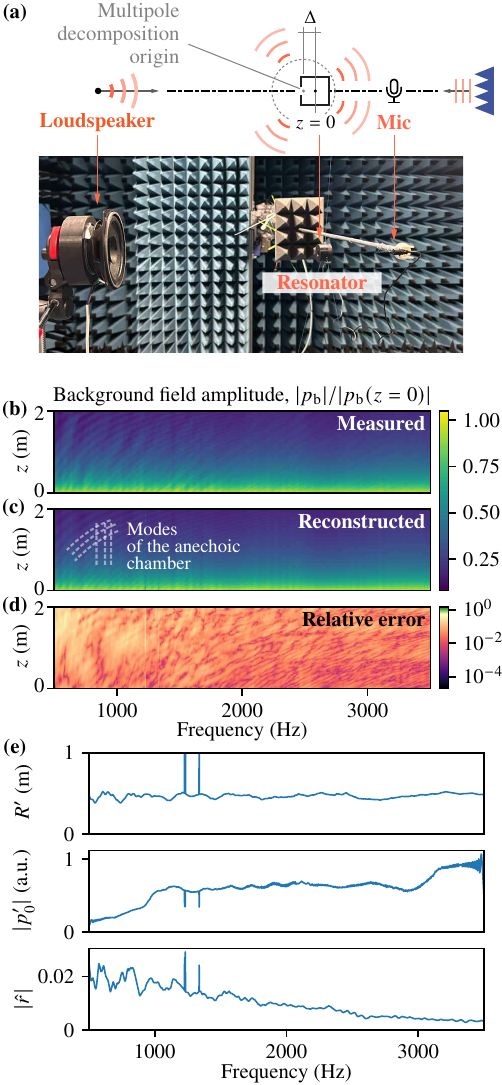}
    \caption{
        \label{fig:model_fit}
        (a) Geometry of the experimental setup placed in the anechoic chamber.
        Microphone was fixed at the mobile stage.
        (b) Amplitude of measured incident wave field $|p_\text{b}^\text{meas}|$, (c) fitted model $|p_\text{b}|$, and (d) the relative error.
        The fields are normalized with $|p_\text{b}^\text{meas}(z=0)|$ to remove frequency response of the microphone and loudspeaker in the figure.
        (e) The fitted parameters of the effective distance to the  loudspeaker $R'$, amplitude of the background field $|p_0'|$, and the reflection coefficient of the walls $|\hat r|$.
    }
\end{figure}

To overcome this problem, we suggest a two-step approach to filter out the parasitic noise coming from the residual rescattering: i) {\it at the first step}, we try to reconstruct the background field with the help of a numerical model, and ii) {\it at the second step}, we fit the measured field scattered from the Helmholtz resonator to extract the ECS.

\begin{figure*}
    \includegraphics{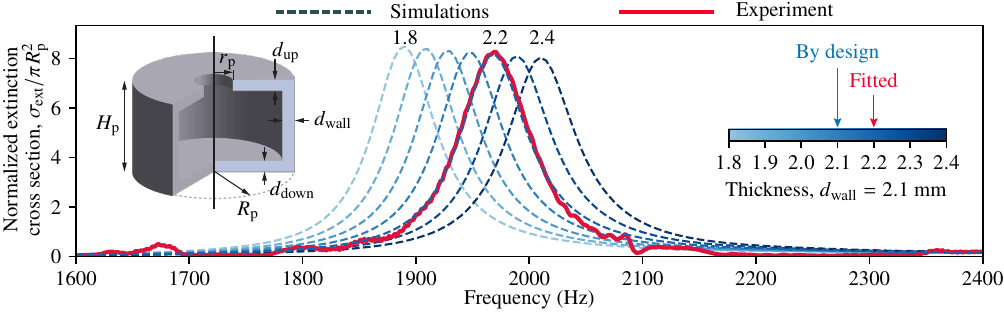}
    \caption{
        \label{fig:main_result}
        The normalized extinction cross section of the Helmholtz resonator found by the two-step approach from experimental data.
        The normalized extinction cross section calculated in simulations for $d_\text{wall}=2.1$ mm (as the resonator was printed on a 3D printer) and also $d_\text{wall}=2.2$ mm to highlight that the resonance shift can be explained as geometric imperfection.
        The geometry of the resonator is shown in the inset.
    }
\end{figure*}

\subsection{Reconstruction of the background field}

At the first step, we have to correct the background field measured without the sample. We utilize the following model of the measured background field
\begin{equation}
    p_\text{b}(z, \omega) = p_0' \left(R' \frac{e^{ikz}}{z + R'} + \hat{r} e^{ik(L-z)}\right).
    \label{eq:p_b_measured}
\end{equation}
Since the exact position of the loudspeaker and the effective center of the scatterer cannot be identified, an additional corresponding phase and amplitude should be added
\begin{equation*}
    R'= R+\Delta,\quad
    p_0'= \frac{R}{R'}p_0 e^{ik\Delta}.
\end{equation*}
The second term in \autoref{eq:p_b_measured} is related to the contribution from the reflected waves, $\hat{r}(\omega)$ is the effective reflection coefficient, and $L$ is the effective distance to the anechoic chamber wall (see \autoref{fig:lines}~(c)).
The physical justification for this effective plane-wave reflection is discussed in \autoref{app:background_field_model}.
At this stage, we need to find the model parameters to match the background field with account for rescattering, which implies solving the  optimization problem
\begin{equation}
    \mathop{{}\operatorname{min}}\limits_{\substack{c\in\mathbb{R}^+,\,p_0'(\omega)\in\mathbb{C}, \\ L\in\mathbb{R}^+,\,R'(\omega)\in\mathbb{R}^+, \\ \hat{r}(\omega)\in\mathbb{C}}} \sum_{i,j}|{p_\text{b}}^{\text{meas}}(z_i, \omega_j)-p_\text{b}(z_i, \omega_j)|^2.
    \label{eq:optimization_problem_1}
\end{equation}
From this, one can retrieve the sound velocity $c$ assumed to be independent of frequency, the amplitude of the emitted wave $p_0'(\omega)$, the effective distance $L$ to the front wall and effective $\hat{r}(\omega)$ reflection coefficient for this wall, and the effective position of the loudspeaker $R'(\omega)$.
The total number of real-valued parameters to optimize is $5N_\omega+2$.

We conducted the experiments in a microwave anechoic chamber (see \autoref{fig:model_fit}~(a)).
It has acoustic anechoic properties significantly improved compared to a usual room, and the reverberation time is below 100 ms for frequencies higher than 300 Hz.
We used a chirp signal with frequencies spanning from 500 Hz to 3500 Hz to generate the incident wave with the loudspeaker.
The sample rate was fixed at 44100 Hz during the recording.
Then we synchronized all recordings using a loopback channel as the reference and applied discrete Fourier transform.
From the resulting spectrum, we selected $N_\omega=1784$ frequency points $\omega_j$.
We scanned the acoustic field along the $z$-axis at positions $z_i$ starting from the future location of backside of the scatterer over a distance of 2~m with a step size of 2~mm, and collected the data using a microphone.

The spectral measurements of the background field [see \autoref{fig:model_fit}~(b)] show distinct ripples in the amplitude signal.
We ran the model optimization in order to reconstruct this field and obtained quite good agreement with relatively low error as shown in \autoref{fig:model_fit}~(c) and \autoref{fig:model_fit}~(d), respectively.
To fit the data with the model parameters, deep learning approaches and procedures were used, such as automatic differentiation tools provided by the PyTorch library~\cite{Ansel_PyTorch_2_Faster_2024}.
Details of the optimization procedure are provided in \autoref{app:optimization}.
The fitting parameters showing reasonable agreement with the predicted values are plotted in \autoref{fig:model_fit}~(e).
For $R'$, the obtained values are close to constant, as the loudspeaker does not move in space.
The found speed of sound $c$ is 349.4 m/s.
The obtained $|p_0'|$ is the amplitude of the response function of the microphone and loudspeaker.
The effective reflection coefficient $|\hat{r}|$ reaches the value of 0.02, which together with the relation $|\Delta p_\text{b}| \approx |\hat{r}||p_0|$ is even larger than the amplitude of the residual rescattering taken in \autoref{fig:noisy_sigma}.
The abrupt spikes in the fitted data around the frequency 1250 Hz are related to the optimization algorithm getting stuck in local minima.

\subsection{The acoustic scattering from a Helmholtz resonator}

At the second step, we reconstruct the ECS of the scatterer accounting for the additional contributions to the scattered field from the residual rescattering.
Since we collect the scattered signal right behind the scatterer where the near-field contributions may be strong, we use the scattered field expansion near the effective center $\Delta$ of the scatterer up to the second order following \autoref{eq:p_s_therory_expansion}:
\begin{equation*}
    p_\text{s}(z-\Delta, \omega) = p_0\left( \frac{f}{kz} + \frac{f_{\text{nf}}}{(kz)^2}\right)e^{ikz} + \Delta p_\text{s},
    \label{eq:p_s_model}
\end{equation*}
where $\Delta p_s$ is the additional unknown contribution to the scattered field.

In order to extract the far-field component, we scan the acoustic field along the $z$-axis at the same $z_i$ coordinates used in the experiment without the sample, and then by subtracting the previously measured background field $p_b^\text{meas}$, we retrieve the scattered part $p_s^\text{meas}$.
Then, we extract the real-valued extinction cross section from the complex-valued scattered pressure.
For that, we introduce the function $\Sigma$
\begin{equation}
    {\Sigma(z, \omega)} =\dfrac{4\pi}{k^2} \operatorname{Im}\left[\frac{{p}_\text{s}(z, \omega)}{p_0'(\omega) e^{ikz}}\right],
    \label{eq:Sigma_def}
\end{equation}
which has the form similar to  the extinction cross section but includes other non-related terms:
\begin{multline*}
    {\Sigma(z-\Delta, \omega)} \\ = \dfrac{R'}{(R'-\Delta) }\frac{1}{kz}\left[{\sigma_\text{ext}} + \frac{4\pi}{k^2} \frac{\operatorname{Im} f_{\text{nf}}(\omega)}{kz} +     {\Delta \Sigma}(\omega)\right],
\end{multline*}
where $\Delta \Sigma(\omega)$ corresponds to the unknown part of the scattered field $\Delta p_\text{s}$, and we assume that this quantity depends only on frequency.
By solving  another optimization problem
\begin{equation}
    \mathop{{}\operatorname{min}}\limits_{\substack{\Delta\in\mathbb{R}^+,\, \sigma_\text{ext}(\omega)\in\mathbb{R}^+, \\ \operatorname{Im} f_{\text{nf}}(\omega)\in\mathbb{R}, \, \Delta \Sigma(\omega)\in\mathbb{R}}} \sum_{i,j}|\Sigma^{\text{meas}}(z_i, \omega_j)-\Sigma(z_i, \omega_j)|^2,
    \label{eq:optimization_problem_2}
\end{equation}
we obtain the function of interest $\sigma_\text{ext}(\omega)$.
The $\Sigma^{\text{meas}}$ is calculated using \autoref{eq:Sigma_def} with substituted measured field and $p_0'(\omega)$ and $c$ found in the first step.
The last problem requires optimizing $3N_\omega + 1$ parameters.

We applied this procedure to measure the ECS of a Helmholtz resonator with the geometry shown in the inset of \autoref{fig:main_result}.
The resonator was printed from PLA plastic material, having the following geometrical parameters: $H_\text{p}=15.1$~mm, $R_\text{p}=11.95$~mm, $r_\text{p}=2.9$~mm, $d_\text{up}=1.9$~mm, $d_\text{down}=d_\text{wall}=2.1$~mm.

Details of the optimization procedure are provided in \autoref{app:optimization}.
The found $\Delta$ is 1.3 cm.
The found ECS is shown in \autoref{fig:main_result} by the red solid line along with the simulated curves obtained from numerical simulations in COMSOL Multiphysics.
For the simulations we used thermoviscous acoustics in the spherical region close to the scatterer, hard wall boundary conditions on scatterer boundaries, and pressure acoustics in the outer domain.
The fitted speed of sound is higher than the nominal value at room temperature and, if interpreted solely in terms of temperature, would correspond to approximately $30^\circ$~C.
The temperature measured before the experiment was approximately $22^\circ$~C, while the temperature and humidity were not continuously monitored during the measurements.
Therefore, the fitted value of $c$ should not be regarded as a direct measurement of the actual thermodynamic conditions in the chamber.
For the numerical simulations, we used the built-in air material parameters for ambient temperature of 293.15~K, but fixed its speed of sound to 343.2~m/s and density to 1.2044 kg/m$^3$.
One can see that both simulations and experimental data demonstrate that the Helmholtz resonator has a single resonance frequency that depends on its geometric parameters~\cite{Ingard1953, Langfeldt2022Oct}.
The found resonance frequency is shifted by 20 Hz (1\% of the absolute frequency).
This discrepancy may originate from both uncertainty in the actual acoustic properties of air during the experiment and fabrication imperfections and inaccuracy.
By varying the wall thickness parameter within 0.1 mm (5\% of the wall thickness), we obtained not only matching of the resonant frequency but also matching of the maximal value and shape of the ECS spectrum.
The obtained result clearly demonstrates that the proposed model-based measurement procedure provides accurate retrieval of the extinction spectrum.
This is also clearly seen in comparison with the ECS spectra obtained without denoising the background and scattering signal and removing the residual rescattering field, shown in \autoref{app:direct_meas}.

In this work, we focus on errors associated with finite source ($R$) and detector ($z$) distances and residual rescattering, while angular and lateral misalignment are not the primary focus.
The angular dependence of the forward-scattered field and its use for alignment were investigated in Ref.~\citen{Smith1985Jan}, where variations in the amplitude and phase around the forward direction were used to align the scatterer and receiver.
Similarly, in Ref.~\citen{Kotelnikova2025Feb}, the acoustic pressure was measured at neighboring spatial points to locate the focus of the incident beam and accurately position the scatterer.
In our experiment, the loudspeaker, scatterer, and microphone scan line were mechanically aligned along the same axis using the moving stage.
The effect of an angular deviation from the exact forward direction for the particular Helmholtz-resonator configuration considered in our experiment is illustrated in \autoref{app:angular_alignment}.

As a note, we should emphasize that it is very important to keep the environment at the first stage, during background field measurements, and at the second stage, during the scattered field measurements, exactly the same, as even the smallest reconfiguration of the anechoic chamber results in changing the background field and causes the ECS retrieval algorithm to fail.
Also, during the data processing, it turned out that some of the recordings were randomly corrupted due to synchronization failures, such as $N_z = 957$ for the first step and $N_z=841$ for the second step.
In both optimization problems, the number of fitted data points, $N_z\cdot N_\omega$, exceeds the number of free parameters, $5N_\omega+2$ and $3N_\omega+1$ for the first and second steps, respectively.

\section{Conclusion}
We developed a practical method for measuring the acoustic extinction cross section of subwavelength resonators using the optical theorem under realistic, non-ideal laboratory conditions.
By extending the theorem to spherical incident waves and analyzing the impact of finite distances and residual reflections, we identified the main limitations of direct measurements.
To overcome them, we introduced a two-step reconstruction procedure --- powered by deep-learning optimization tools --- to model the background field and compensate for parasitic rescattering.

Applying this approach to a 3D-printed Helmholtz resonator, we obtained extinction spectra in excellent agreement with full-wave simulations and sensitive to small geometric imperfections.
These results demonstrate that forward-scattering measurements, combined with physics-guided deep-learning techniques, provide a reliable and accessible tool for characterizing acoustic resonators.
The methodology offers a reproducible framework that can be extended to broader acoustic scattering and metamaterial applications.

The experimental and simulation data, code with additional examples, and COMSOL model-generation scripts are openly available in Ref.~\citen{Igoshin2026May}.

\begin{acknowledgments}
    We thank Ivan Toftul for the inspiration and permanent help. The work is funded by Russian Science Foundation, grant № 25-79-31027, https://rscf.ru/project/25-79-31027/.
\end{acknowledgments}

\section*{Author Declarations}
\subsection*{Conflict of Interest}
The authors have no conflicts to disclose.

\section*{Data Availability}
The data that support the findings of this study are openly available in
djiboshin/paper-optical-theorem at https://doi.org/10.5281/zenodo.20049963.

\appendix

\section{Expansion of the spherical incident wave}
\label{app:expansion}

We consider a point source located on the negative $z$ axis at a distance $R$ from the origin.
With the normalization $p_0=p_\text{b}(0)$, the incident field is
\begin{equation}
    p_\text{b}(z) = p_0 kR \dfrac{e^{ik(z+R)}}{k(z+R)}e^{-ikR}  = p_0 \dfrac{e^{ik(z+R)}}{k(z+R)} \frac{kR}{e^{ikR}}
    \label{eq:p_b_appendix}
\end{equation}

Introduce $u=kR$, $v=kz$ and $w=u+v$.
For $|v|<|u|$, or equivalently $|z|<R$,
\begin{multline*}
    \dfrac{e^{ik(z+R)}}{k(z+R)} = \dfrac{e^{iw}}{w} = \dfrac{\cos(w)}{w} + i\dfrac{\sin(w)}{w} \\
    =\sum_{\ell=0}^\infty (-1)^\ell(2\ell+1) j_\ell(v) \left[-y_\ell(u) + i j_\ell(u) \right] \\
    = i \sum_{\ell=0}^\infty (-1)^\ell(2\ell+1) j_\ell(v) h_\ell^{(1)}(u) \\
    = \sum_{\ell=0}^\infty i^\ell (2\ell+1) j_\ell(v)  \left[(i)^{\ell+1} h_\ell^{(1)}(u)\right],
\end{multline*}
where $y_\ell$ is the spherical Neumann function and $h_\ell^{(1)}=j_\ell+iy_\ell$.
Substitution into \autoref{eq:p_b_appendix} yields
\begin{equation}
    p_\text{b}(z) = p_0\sum_{\ell=0}^{\infty}i^\ell(2\ell+1)j_\ell(kz)B_\ell(kR),
    \label{eq:p_b_expansion_appendix}
\end{equation}
with
\begin{equation}
    B_\ell(u) = (i)^{\ell+1} \frac{u}{e^{iu}} h_\ell^{(1)}(u) = 1 + \frac{i(\ell+1)\ell}{2u} + \mathcal{O}\left(\frac{1}{u^2}\right).
    \label{eq:hankel_expansion}
\end{equation}
This is the expansion used in \autoref{eq:p_b_expansion}.

The condition $|z|<R$ applies only to the region in which the incident field is expanded.
Because the expansion is required only inside a sphere of radius $a$ enclosing the scatterer, it is sufficient to require $a<R$.
In the experiment, the fitted mean source-to-effective-center distance is $R=R'-\Delta\approx46$~cm.
With the effective center at $z=-\Delta=-1.3$~cm and the back face of the resonator at $z=0$, the radius of a sphere enclosing the resonator is approximately $a=\sqrt{\Delta^2+R_\text{p}^2}\approx1.8~\text{cm}$, so that $a<R$.

As discussed in the main text, the condition $R\gg ka^2$ is required for the validity of expansion~\autoref{eq:hankel_expansion}.
At 2400~Hz, using $c=343.2$~m/s and $a\approx1.8$~cm, one obtains $ka^2\approx1.4$~cm, well below the experimental value $R\approx46$~cm.

\section{Forward-scattering asymptotics for spherical-wave incidence}
\label{app:f_expansion}

At a detector located on the $z$ axis, only the $m=0$ spherical harmonics contribute because
\begin{equation*}
    Y_\ell^m(0,\varphi)=\sqrt{\frac{2\ell+1}{4\pi}}\,\delta_{m0}.
\end{equation*}
The forward scattered field can therefore be written as
\begin{multline}
    p_\text{s}(z) = \sum_{\ell=0}^\infty \sum_{m=-\ell}^{\ell} \mathbf{p}_{\text{s}, \ell m} Y_\ell^m(\theta=0, \varphi=0)h_\ell^{(1)}(kz)\\
    = \sum_{\ell=0}^\infty \sqrt{\frac{2\ell+1}{4\pi }} \mathbf{p}_{\text{s}, \ell,m=0} h_\ell^{(1)}(kz) \\
    = \sum_{\ell=0}^\infty\sum_{\ell'=0}^\infty\sum_{m'=-\ell'}^{\ell'} \sqrt{\frac{2\ell+1}{4\pi }} \mathbf{\hat{T}}_{\ell, m=0}^{\ell' m'} \mathbf{p}_{\text{b}, \ell' m'} h_\ell^{(1)}(kz)
    \label{eq:p_s_theory_expansion}
\end{multline}
Because the point-source field is axially symmetric about the $z$ axis, its incident-wave coefficients are
\begin{equation}
    \mathbf{p}_{\text{b}, \ell' m'} = p_0\sqrt{4\pi(2\ell'+1)}i^{\ell'} \delta_{m',0} B_{\ell'}(kR)
    ,
    \label{eq:p_b_lm_expansion}
\end{equation}
which is related to the expansion of the spherical incident wave in \autoref{eq:p_b_expansion_appendix}.
Then
\begin{equation*}
    p_\text{s}(z) =p_0\frac{e^{ikz}}{kz} \sum_{\ell,\ell'=0}^{\infty} \gamma_{\ell}^{\ell'}i^{\ell'-\ell-1} B_{\ell'}(kR)B_\ell(kz).
\end{equation*}
where $\gamma_{\ell}^{\ell'} =\sqrt{(2\ell+1)(2\ell'+1)}\mathbf{\hat T}_{\ell0}^{\ell'0}$.

For PW incidence and in the far field, $B_{\ell'}(kR)\to1$ and $B_\ell(kz)\to1$, so the forward scattering amplitude is
\begin{equation*}
    f=\sum_{\ell,\ell'=0}^{\infty} \gamma_{\ell}^{\ell'}i^{\ell'-\ell-1}.
\end{equation*}

Keeping the first corrections in $1/(kR)$ and $1/(kz)$ gives
\begin{multline}
    p_\text{s}(z) \\
    = p_0 \frac{e^{ikz}}{kz}\left[f + \frac{F_R}{kR} + \frac{F_z}{kz} + \frac{F_{Rz}}{k^2Rz} + \frac{F_{R^2}}{k^2R^2} + \frac{F_{z^2}}{k^2z^2} + \dots\right] \\
    = p_0 \frac{e^{ikz}}{kz}\left[f + \mathcal{O}\left(\frac{1}{kR} + \frac{1}{kz}\right)\right].
\end{multline}
Thus,
\begin{equation*}
    f =\frac{p_\text{s}(z)}{p_0e^{ikz}}kz + \mathcal{O}\left(\frac{1}{kR}+\frac{1}{kz}\right).
\end{equation*}

\section{Scattering by a sphere}
\label{app:sphere}

For a homogeneous fluid sphere, spherical symmetry makes the T-matrix diagonal,
\begin{equation*}
    \mathbf{\hat{T}}_{\ell m}^{\ell' m'} = a_\ell \delta_{\ell, \ell'} \delta_{m,m'},
\end{equation*}
where $a_\ell$ is the acoustic Mie coefficient~\cite{Toftul2025Dec,Toftul2025Sep},
\begin{equation}
    a_\ell = \frac{\alpha j_\ell'(ka)j_\ell(k'a) - j_\ell(ka)j_\ell'(k'a)}{h_\ell^{(1)}(ka)j_\ell'(k'a) - \alpha h^{(1)\prime}_\ell(ka)j_\ell(k'a)},
\end{equation}
with $\alpha = \frac{\rho_\text{sphere}}{\rho_\text{host}}\cdot\frac{c_\text{sphere}}{c_\text{host}}$, $k' = k\frac{c_\text{host}}{c_\text{sphere}}$.

For SW incidence, \autoref{eq:p_b_lm_expansion} gives the forward scattered field
\begin{multline*}
    p_\text{s}(z) = \sum_{\ell=0}^{\infty} \sqrt{\frac{2\ell+1}{4\pi }} a_\ell \mathbf{p}_{\text{b}, \ell, m=0} h_\ell^{(1)}(kz) \\
    = p_0 \sum_{\ell=0}^{\infty} (2\ell+1) a_\ell i^\ell B_\ell(kR) h_\ell^{(1)}(kz).
\end{multline*}
The corresponding incident field at the detector is evaluated from the exact point-source expression~\autoref{eq:p_b_appendix}.
The colored curves in \autoref{fig:opt_t} are calculated from the leading explicit term of \autoref{eq:opt-theorem-SW},
\begin{equation}
    \sigma_\text{ext}(R,z)
    =\frac{4\pi}{k}\operatorname{Im}\left[
        \frac{p_\text{s}(z)}{p_\text{b}(z)}\right]\frac{zR}{z+R}.
\end{equation}

The finite-$R$ and finite-$z$ error is obtained by comparison with the exact PW result
\begin{equation}
    \sigma_\text{ext} = -\frac{4\pi}{k^2}\sum_{\ell=0}^{\infty} (2\ell+1) \operatorname{Re}(a_\ell).
    \label{eq:sigma_ext_sphere}
\end{equation}

For the numerical calculations, series are truncated at $\ell_\text{max}=60$, which is sufficient over the $ka$ range considered in \autoref{fig:opt_t}.
The code used to reproduce these calculations is available in Ref.~\citen{Igoshin2026May}.

\section{Contribution of residual rescattering and measurement noise to the ECS}
\label{app:sigma_noisy}

The measured extinction cross section has the following form:
\begin{equation*}
    \widetilde{\sigma}_\text{ext} = \frac{4\pi}{k}\operatorname{Im}\left[\frac{p_\text{s}(z) + \Delta p_\text{s}}{p_\text{b}(z) + \Delta p_\text{b}}\right]\frac{zR}{z+R}.
\end{equation*}
Assuming  $|\Delta p_\text{b}| \ll |p_\text{b}|$
\begin{equation*}
    \widetilde{\sigma}_\text{ext} = \frac{4\pi}{k}\operatorname{Im}\left[\left(\frac{p_\text{s}}{p_\text{b}} + \frac{\Delta p_\text{s}}{p_\text{b}}\right)\left(1-\frac{\Delta p_\text{b}}{p_\text{b}}\right)\right]\frac{zR}{z+R}.
\end{equation*}
In the far-field region
\begin{equation*}
    \widetilde{\sigma}_\text{ext} \approx \frac{4\pi}{k}\operatorname{Im}\left[\left(\frac{f}{k} + \frac{\Delta p_\text{s}}{p_0 e^{ikz}}z\right)\left(1-\frac{\Delta p_\text{b}}{p_0e^{ikz}}\frac{z+R}{R}\right)\right].
\end{equation*}
Expanding brackets and dropping $\Delta p_\text{b}\cdot\Delta p_\text{s}$ term
\begin{equation*}
    \widetilde{\sigma}_\text{ext} \approx \sigma_\text{ext} +\frac{4\pi}{k}\operatorname{Im}\left[\frac{\Delta p_\text{s}}{p_0 e^{ikz}}z - \frac{f}{k}\frac{\Delta p_\text{b}}{p_0e^{ikz}}\frac{z+R}{R}\right].
\end{equation*}

\section{Direct measurements}
\label{app:direct_meas}

\begin{figure}
    \includegraphics{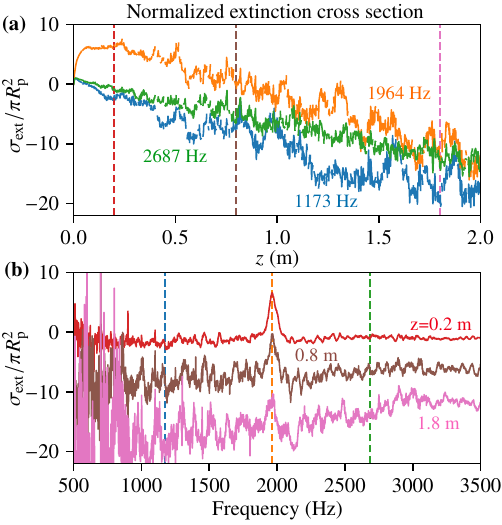}
    \caption{
        \label{fig:direct_meas}
        (a) Normalized extinction cross section at different frequencies, indicated by different colors.
        (b) Normalized extinction cross section calculated at different $z$ positions.
        As $z$ increases, the resonance becomes less distinguishable due to the background level from rescattered waves.
        The ECS also becomes negative.
        The plots and vertical lines in (a) and (b) correspond via matching colors.
    }
\end{figure}

To demonstrate how residual rescattering contributes to the results, the ECS calculated using \autoref{eq:opt-theorem-SW} is shown in \autoref{fig:direct_meas}.
As seen in \autoref{fig:direct_meas}~(a), the ECS decreases linearly with increasing $z$.
This linear trend is connected to the linear error growth discussed in the main text (see the comments on \autoref{eq:sigma_noisy_error}).

\section{Measured background field model}
\label{app:background_field_model}

\begin{figure}
    \includegraphics{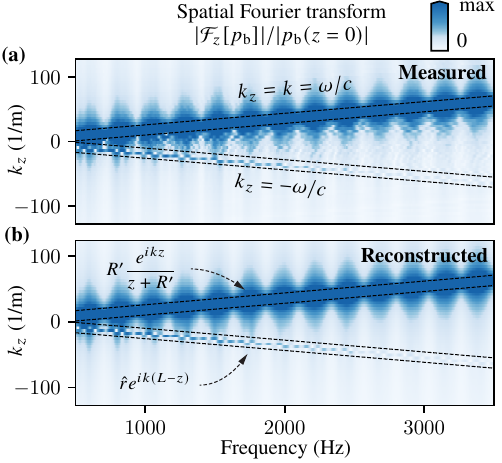}
    \caption{
    \label{fig:background_field_model}
    Spatial spectrum of the background field.
    Absolute value of the spatial Fourier transform of the normalized complex background pressure field, $|\mathcal{F}_z[p_\text{b}(z,\omega)]|/|p_\text{b}(0,\omega)|$ for (a) the measured and (b) reconstructed background fields.
    The forward-propagating spherical-wave contribution is centered around $k_z=\omega/c$, whereas the narrower branch near $k_z=-\omega/c$ corresponds to the reflected field represented by the effective plane-wave term in \autoref{eq:p_b_measured}.
    The reference dispersion lines $k_z=\pm\omega/c$ are calculated using $c=349.4$~m/s.
    }
\end{figure}

For each frequency, we calculate the spatial Fourier transform of the complex pressure field along the measurement axis,
\begin{equation*}
    p_\text{b}(k_z,\omega)
    = \mathcal{F}_z[p_\text{b}(z, \omega)].
\end{equation*}
The absolute value of the resulting spatial spectrum is shown in \autoref{fig:background_field_model}.
Only the region $z>0.5$~m is used.
Excluding the region close to the scatterer position reduces the spatial variation of the spherical-wave amplitude, proportional to $1/(z+R)$, and allows the weaker reflected contribution to be distinguished more clearly in the spatial spectrum.

\autoref{fig:background_field_model}~(a) shows the spatial spectrum obtained directly from the measured background field.
Two distinct branches are observed.
The broader branch follows $k_z = \omega / c$ and corresponds to the forward-propagating field emitted by the loudspeaker.
Although its phase along the measurement axis varies as $e^{ikz}$, its amplitude varies as $1/(z+R)$, thus its Fourier spectrum is broadened around $k_z=\omega/c$.

In addition, a considerably narrower branch is observed near $k_z = -\omega / c$.
This branch corresponds to a wave propagating in the negative $z$ direction.
This experimentally observed component therefore motivates its representation in \autoref{eq:p_b_measured} by the effective plane-wave term
\begin{equation*}
    p_{\mathrm{r}}(z,\omega) = p_0'\hat r(\omega)e^{ik(L-z)}.
\end{equation*}

\autoref{fig:background_field_model}~(b) shows the same spatial Fourier analysis applied to the fitted complex background field $p_\text{b}$.
The reconstruction reproduces both the forward branch near $k_z=\omega/c$ and the narrow backward branch near $k_z=-\omega/c$, confirming that the spherical-wave plus effective plane-wave model captures the dominant components of the experimentally measured field.

\section{Optimization details}
\label{app:optimization}

Both inverse problems, \autoref{eq:optimization_problem_1} and \autoref{eq:optimization_problem_2}, were solved with the Adam optimizer implemented in PyTorch.
We used the standard Adam parameters $\beta_1=0.9$, $\beta_2=0.999$, and $\epsilon=10^{-8}$.
Random mini-batches of the measured $z$ positions were used during optimization to help the optimizer avoid poor local minima.

For the background-field reconstruction in \autoref{eq:optimization_problem_1}, the learning rate was 0.1 and the mini-batch size was 200.
The parameters were initialized as $c\approx400$~m/s, $L\approx7.3$~m, $R'(\omega)\approx1.99$~m, $p_0'(\omega)\approx1.3$~a.u., and $\hat r(\omega)\approx0.049$.

For the ECS reconstruction in \autoref{eq:optimization_problem_2}, the learning rate was 0.01 and the mini-batch size was 350.
The parameters were initialized as $\Delta\approx0.006$~m, $\sigma_\text{ext}(\omega)\approx0.007$~m$^2$, $\operatorname{Im}f_\text{nf}(\omega)=0$, and $\Delta\Sigma(\omega)=0$~m$^2$.

The first and second optimization problems were run for fixed numbers of 5000 and 8000 epochs, respectively.
No early-stopping criterion was used.
In both cases, convergence was checked by verifying that the objective function had reached a plateau by the end of the optimization.

The code used to reproduce the optimization procedure and the reported results is available in Ref.~\citen{Igoshin2026May}.

\section{Angular alignment}
\label{app:angular_alignment}

\begin{figure}
    \includegraphics{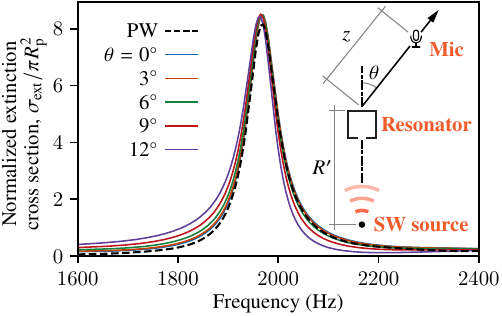}
    \caption{
        Effect of angular misalignment on the retrieved extinction cross section.
        The normalized ECS obtained from the optical theorem \autoref{eq:opt-theorem-SW} is shown for different observation angles $\theta$ at $z=2$~m.
        The dashed black curve shows the reference plane-wave ECS.
        The calculations are performed for the Helmholtz resonator with $d_\text{wall}=2.2$~mm and $R'=47$~cm.
        The inset illustrates the geometry and definition of the angle $\theta$.
        \label{fig:angular_alignment}
    }
\end{figure}

To estimate the sensitivity of the retrieved ECS to imperfect angular alignment, we numerically displaced the observation direction from the exact forward direction by an angle $\theta$, while keeping the spherical-wave source and the scatterer aligned along the $z$ axis (see \autoref{fig:angular_alignment}).
The microphone was placed at a distance $z=2$~m from the back side of the scatterer.
The calculations were performed for the Helmholtz resonator considered in the main text, with $d_\text{wall}=2.2$~mm and a source distance $R'=47$~cm, corresponding to the mean value obtained from the experimental reconstruction.
\autoref{fig:angular_alignment} compares the ECS obtained directly from the leading explicit term of \autoref{eq:opt-theorem-SW} at different observation angles with the reference plane-wave ECS.
The deviation increases with $\theta$, as expected because the optical theorem formally requires measurement in the exact forward direction, but remains relatively small over the considered angular range.

\bibliography{main.bib}

\end{document}